\documentclass[11pt]{article}
\usepackage{epstopdf}
\usepackage[a4paper,hmarginratio=1:1,vmarginratio=2:3,totalwidth=15.3cm,totalheight=22.7cm]{geometry}
\usepackage{bm,epstopdf,epsfig,amsmath,amssymb,amsfonts,colordvi,latexsym,comment,cancel,
verbatim,slashed}
 \usepackage{graphics}
\usepackage[font=md,captionskip=8pt]{subfig}
\usepackage[usenames,dvipsnames]{color}

\usepackage{setspace}
\usepackage[noadjust]{cite}

\usepackage[utf8]{inputenc}

\newcommand{\cO}{{\cal O}}

\newcommand{\Li}{\mbox{Li}}
\newcommand{\Tr}{\mbox{Tr}}

\newcommand{\ra}{\rightarrow}
\newcommand{\be}{\begin{equation}}
\newcommand{\ee}{\end{equation}}
\newcommand{\bea}{\begin{eqnarray}}
\newcommand{\eea}{\end{eqnarray}}
\newcommand{\Ra}{\Rightarrow}

\newcommand{\baa}{\begin{array}}
\newcommand{\eaa}{\end{array}}
\long\def\symbolfootnote[#1]#2{\begingroup
\def\thefootnote{\fnsymbol{footnote}}\footnote[#1]{#2}\endgroup}
\begin{document} 
\begin{flushright}
\end{flushright}

\bigskip\medskip

\thispagestyle{empty}

\vspace{3.5cm}

\begin{center}
  {\Large {\bf Quantum implications  of a scale invariant regularisation}}

\vspace{1cm}

 {\bf D. M. Ghilencea}
\symbolfootnote[1]{E-mail: dumitru.ghilencea@cern.ch}

\bigskip

 {\small Theoretical Physics Department, National Institute of Physics}

{\small and Nuclear Engineering Bucharest, 077125 Romania}

\end{center}

\begin{abstract}
\noindent
We study scale invariance at the quantum level in a perturbative approach. 
For  a  scale-invariant classical theory the scalar potential is computed  
at three-loop level while keeping manifest this symmetry.
Spontaneous scale symmetry breaking is transmitted at quantum level
to the visible sector (of $\phi$) by the  associated Goldstone mode (dilaton $\sigma$)
which enables a scale-invariant regularisation and  whose vev $\langle\sigma\rangle$
generates the subtraction scale ($\mu$). While the hidden ($\sigma$) and visible sector 
($\phi$) are classically decoupled in $d\!=\!4$ due to an enhanced Poincar\'e symmetry, 
they interact through (a  series of) evanescent couplings  $\propto\epsilon^k$, ($k\!\geq\! 1$), 
dictated  by the  scale invariance  of the action in  $d\!=\!4-2\epsilon$. 
At the  quantum level these couplings generate new corrections to the potential, 
such as scale-invariant  non-polynomial effective operators $\phi^{2n+4}/\sigma^{2n}$
and also log-like terms ($\propto \ln^k \sigma$) restoring the scale-invariance of
known quantum corrections.
The former are comparable in size to ``standard'' loop corrections and
important for  values of $\phi$ close to $\langle\sigma\rangle$.
 For $n=1,2$ the beta functions of their coefficient are computed 
at three-loops. In the infrared (IR)  limit the dilaton fluctuations decouple,  the effective 
operators are suppressed by large $\langle\sigma\rangle$ and the effective potential 
becomes that  of  a renormalizable  theory with explicit scale symmetry breaking by 
the ``usual'' DR scheme (of $\mu=$constant).
\end{abstract}

\newpage
\section{Introduction}

It is a common view  that Standard Model (SM)  is only a  low-energy effective
 theory and  ``new physics'' could  arise at some scale  below $M_{\rm Planck}$.
 The scale of ``new physics'' can be the vacuum expectation value (vev) of a scalar field
$\sigma$ present beyond the SM spectrum.  It is then natural to ask
how the higgs  mass is  protected from large  quantum corrections   
associated with  $\langle\sigma\rangle$\footnote{In the absence of 
``new physics'' below $M_{\rm Planck}$ and ignoring gravity,
SM has no hierarchy problem.}. One long-held answer is TeV-supersymmetry\footnote{
TeV-SUSY models have large fine-tuning \cite{G1} which cannot coexist with
a good data fit  ${\chi^2}/{\rm dof}\!\approx\! 1$ \cite{G2}.}.

Scale invariance  may also protect the higgs mass against large 
quantum corrections. This starts from the observation 
that for  a vanishing higgs mass parameter, SM has an increased symmetry: it is
scale invariant. This means that  the classical action  is invariant 
under a transformation: $x\!\ra\! \rho^{-1} x,\,\, \phi\!\ra\! \rho^{d_\phi}\phi$\,
\,($d_\phi$ is the mass dimension of $\phi$). 
Scale symmetry was noticed to play a role in protecting the electroweak
scale \cite{M1,F1,Foot:2007iy,I1}
with classically scale-invariant extensions of the SM considered in
\cite{F2,A1,Hur:2011sv,
Ishiwata:2011aa,Okada:2012sg,Iso:2012jn,Englert:2013gz,Heikinheimo:2013fta,
Heikinheimo:2013xua,Hambye:2013sna,Bars:2013yba,Carone:2013wla,Khoze:2013uia,Gorbunov:2013dqa,GHK,
Allison:2014zya,KT,Allison:2014hna,GE,KKE}.
But to address the mass hierarchy problem one must go beyond the classical 
scale symmetry, since the counterterms are actually dictated by the quantum symmetry.
This could protect naturally \cite{thooft}  the higgs mass from large quantum  
corrections \cite{Bardeen} associated with a high scale $\langle\sigma\rangle$
 of ``new physics''. For studies of quantum scale invariance 
(broken spontaneously) and applications to SM  see  
\cite{Englert,S1,S2,S3,tamarit,dg1,Ghilen,dzp,Monin,Ferreira}.

Our goal is to  study further the  models in which the classical scale symmetry 
is extended at the  quantum level and is broken only spontaneously\footnote{
Classical scale symmetry is often broken by quantum calculations since the UV  regulator 
breaks  it {\it explicitly}. The classical flat direction is then lifted and a 
light pseudo-Goldstone  boson exists. See for example \cite{Foot:2007iy}. 
We do not follow this approach and implement instead a quantum scale symmetry.}.
In such  theory  all scales are generated by fields' vev's.
Such theory can predict ratios of scales (vev's) only, in terms of ratios of 
{\it dimensionless} couplings. A hierarchy of physical mass scales  can then be generated by 
a hierarchy of such couplings. The latter is easier to protect by a 
symmetry (e.g. an enhanced  Poincar\'e symmetry \cite{Foot}) and is  more
 fundamental than a hierarchy of  dimensionful physical scales.
Indeed, in a fundamental theory, any physical scale   should ultimately  be 
determined in terms  of dimensionless  couplings and fields vev's.

Since scale symmetry is broken in the real world, we assume  it 
is broken spontaneously. A flat direction exists and the spectrum contains
the associated Goldstone boson (dilaton, hereafter  $\sigma$) beyond the spectrum 
of the initial model. The subtraction scale  $\mu$ (used in  loop calculations) 
that would  break quantum scale symmetry {\it explicitly},   is replaced by the field 
$\sigma$  which thus  maintains scale symmetry at quantum level and after  
spontaneous breaking generates $\mu\!\sim\! \langle\sigma\rangle$, see Englert et al  
\cite{Englert}. This gives a  scale-invariant regularisation (SR)\footnote{
Versions  of this scheme were used in \cite{S1,S2,S3,tamarit,dg1,Ghilen,dzp,Monin}
(in some cases classical higgs-dilaton mixing was present which concealed the
enhanced Poincar\'e symmetry and the effects discussed below).}.

In this paper we discuss further consequences of the original idea of Englert
 et al \cite{Englert}.  The SR  scheme can be  applied to  any gauge theory, 
although we restrict  our study to a scalar theory. We study more quantum effects
in this scheme  and stress the role of symmetries. In $d=4$ the hidden sector (of 
the dilaton  $\sigma$) is classically decoupled  from the visible sector
 (of higgs-like $\phi$), by invoking  an enhanced  Poincar\'e symmetry   
$P_v\times P_h$ of these two sectors \cite{Foot}.  At the quantum level, the manifest  
scale symmetry   of the action  in  $d=4-2\epsilon$ introduces  evanescent couplings 
$\propto\epsilon\, \tilde\sigma/\langle\sigma\rangle$  of the hidden 
to  the visible sector\footnote{By evanescent coupling we understand  a coupling 
that is non-zero in $d=4-2\epsilon$ and is vanishing in $d=4$.} 
($\tilde\sigma$:  dilaton fluctuations). The  SR scheme  is thus reformulated 
as an ``ordinary'' DR scheme of $\mu$=constant ($\propto\!\!\langle\sigma\rangle$)
plus an additional field ($\sigma$) with  an {\it infinite series} of evanescent  
couplings to the visible sector.

At the  quantum level, such evanescent couplings 
have physical effects. When these couplings multiply  poles of momentum integrals, 
they generate  new (finite or infinite) counterterms, all scale invariant. 
For example one finds   non-polynomial operators  generated
radiatively, such as $\phi^{2n+4}/\sigma^{2n}$, $n\geq 1$ (but also higher 
derivative operators  suppressed by  $\sigma$). They can  transmit scale symmetry 
breaking to   the visible sector.  Such operators can be understood via their Taylor expansion about 
$\sigma=\langle\sigma\rangle+\tilde\sigma$, when they become polynomial.  Scale symmetry acts at the
quantum level as an organising principle that  re-sums the polynomial ones.
We shall study closer these operators, since they are important
at large $\phi$.  Because of their presence, the quantum scale invariant theory is 
non-renormalizable.

We compute  in a manifest scale invariant way the quantum 
corrections to  the scalar potential  in  two-loop  order (diagrammatically)
and three-loop (via Callan-Symanzik equation), for a scale-invariant classical theory. 
The two-loop (three-loop)  potential contains effective  operators as finite (infinite) 
counterterms, respectively.  In the infrared (IR) decoupling limit of the dilaton  
(large $\langle\sigma\rangle$) effective operators vanish;  one then recovers 
the effective potential and trace anomaly of a renormalizable theory (if classical 
theory was so)  with only classical scale  invariance and {\it explicit} scale symmetry 
breaking (SSB) by the ``usual'' DR scheme of $\mu=$constant (no dilaton). The combined  role of quantum 
scale invariance and enhanced  Poincar\'e symmetry in protecting the scalar  mass
at large $\langle\sigma\rangle$ is also reviewed.

Since $M_{\rm Planck}$ breaks scale symmetry, this analysis is valid  for field 
values well below this scale.  One should  extend this study to a  
Brans-Dicke-Jordan theory of gravity with non-minimal coupling where  the dilaton 
vev $\langle\sigma\rangle$  fixes spontaneously $M_{\rm Planck}$. We restrict the analysis to a 
perturbative (quantum)  scale symmetry. At very high momentum scales   some couplings 
(e.g. hypercharge) may  become non-perturbative, but such scale is above $M_{\rm Planck}$, 
where flat  space-time description used here fails anyway.

\section{From classical to quantum scale invariance}
\subsection{Implementing quantum scale invariance}\label{sub1}

Consider a classical scale invariant action, e.g. 
a toy model or the SM with vanishing higgs mass parameter, extended  by 
the dilaton $\sigma$.   We assume that  there is no classical interaction 
between the  visible sector (of fields $\phi_j$)  and the hidden  
sector (of dilaton $\sigma$).  Then
\bea
S=\int d^4x\, L_v(\phi_j,\partial\phi_j)+\int d^4 y \,L_h(\sigma,\partial\sigma)
\eea
The action in $d\!=\!4$ has an  enhanced Poincar\'e symmetry ($P_v\!\times\! P_h$) 
associated with both sectors, which  forbids  a classical coupling $\lambda_m\,\phi_j^2\sigma^2$. 
Such coupling can be naturally  set to $\lambda_m\!=\!0$
and remains so at the quantum  level\footnote{Technically 
$\beta_{\lambda_m}\propto\lambda_m$ at two-loop  \cite{dzp}.}  ``protected''
by this  symmetry \cite{Foot}.

Below  we work with the canonical dilaton $\sigma$  related to the actual Goldstone 
by  $\sigma=\langle\sigma\rangle e^\tau$, so that it  transforms in a ``standard'' way 
under scaling while $\tau$ transforms with a shift
\bea
x\ra \rho^{-1}\,x, \qquad  \sigma\ra \rho\, \sigma,\qquad
\tau\ra \tau + \ln\rho
\eea
The most general potential for $\sigma$ allowed by scale invariance in $d=4$ is then
$\kappa_0 e^{4\tau}\sim \lambda_\sigma \sigma^4$. But Poincar\'e symmetry in the 
dilaton sector demands a flat potential, so $\lambda_\sigma=0$ \cite{Fubini}.
Demanding spontaneous scale symmetry  breaking  $\langle\sigma\rangle\not=0$
means ``we live'' along a  flat direction. This is in the end a tuning 
of the cosmological  constant  and is present anyway in e.g. TeV supersymmetry.
The details of how $\sigma$ acquires  a vev are not relevant below.

At the quantum level it is natural to use the dilaton to generate dynamically
the subtraction scale $\propto\!\langle\sigma\rangle$ in order to preserve scale 
symmetry during quantum  calculations  \cite{Englert}.  We use DR in  
$d=4-2\epsilon$, then   the only possibility  dictated by dimensional 
arguments\footnote{$\mu$  has mass dimension one, while $\sigma$ and 
 $\langle\sigma\rangle$ have dimension $(d-2)/2$.} is
\medskip
\bea\label{mu}
\mu=z\,\sigma^{2/(d-2)}
\eea

\medskip\noindent
with $z$ is an arbitrary  dimensionless parameter (scaling factor);
it  keeps track of the vev of  $\sigma$ after SSB.
The $d\!=\!4$ potential  $V(\phi_j)$ of the visible sector is then analytically continued 
to $d=4-2\epsilon$, into $\mu^{2\epsilon} V(\phi_j)$. Therefore the 
potential in $d=4-2\epsilon$ is actually
\medskip
\bea\label{tildeV}
\tilde V(\phi_j,\sigma)= \Big[z\, \sigma^{2/(d-2)}\Big]^{4-d}\, V(\phi_j),
\eea

\medskip\noindent
and becomes a function of $\sigma$! This ensures the $d=4$ couplings 
remain dimensionless in $d=4-2\epsilon$ and can be used for perturbative calculations.
Therefore, the visible ($\phi_j$) and hidden ($\sigma$)  sectors have evanescent
couplings dictated by the scale symmetry alone of the 
(regularized) action in $d=4-2\epsilon$.  To see these couplings expand (\ref{tildeV}) 
 in powers of $\epsilon$ (loops) and then in terms of fluctuations 
$\tilde\sigma$ about the vev $\langle\sigma\rangle$  of $\sigma$: 
\medskip
\bea\label{tv}
\!\!\!\!
\tilde V(\phi_j,\sigma)=
\mu_0^{2\epsilon}
\Big[ 1\!\!\!\!&+&\!\!\!\!
2 \epsilon\,\Big( \eta- \frac12 \eta^2 +\frac13 \eta^3+\cO(\eta^4)\Big)
\nonumber\\
&+&\!\!\!\!\epsilon^2 \,\Big(2 \eta+\eta^2 -\frac43 \eta^3+\cO(\eta^4)\Big)
+\cO(\epsilon^3)\Big]\, V(\phi_j)
\eea
where
\bea
\mu_0=z\langle\sigma\rangle^{\frac{1}{1-\epsilon}},\qquad
\sigma=\langle\sigma\rangle+\tilde \sigma,\qquad  
\eta=\frac{\tilde\sigma}{\langle\sigma\rangle}.
\eea

\medskip\noindent
A scale invariant regularization  is then re-expressed as an ordinary DR scheme with $\mu=\mu_0$ 
plus an extra field ($\sigma$) with {\it (infinitely many)} evanescent couplings eq.(\ref{tv}).
Since the lhs is scale invariant, so is the rhs if one does not truncate
the expansion in field fluctuations. In practice one can still use a
truncated expansion (see below). From eq.(\ref{tv}) one can read the new
vertices of  evanescent interactions $\propto\!\epsilon^n$ ($n\!\geq\! 1$),  
between $\tilde \sigma$ and $\phi_j$ and the Feynman rules of the scale 
invariant quantum theory\footnote{Field-dependent masses and
propagators also  acquire $\epsilon$ shifts, relevant at loop level.}. 
While these  interactions   vanish  in $d\!=\!4$ or in the dilaton decoupling
 limit ($\eta\ra 0$), at the loop level  have  physical effects.

At quantum level, a coupling proportional to 
 $\epsilon^n$, ($n\geq 1$) in an  amplitude  can bring  new  corrections
to it when multiplying the poles $1/\epsilon^k$ of the  integrals over loop momenta.
One generates finite quantum corrections 
(if $n\!=\!k$) or  new poles/counterterms ($n\!<\!k$) beyond those of the theory 
with $\mu=$constant. If $n\!=\!k$, a scattering  amplitude that involves 
the dilaton depends only on the couplings  of initial $d\!=\!4$  theory,  
without new parameters needed (counterterm couplings).   This can be used to set 
strong lower bounds on  the scale  $\langle\sigma\rangle$.

Since the new  couplings are suppressed, $\eta\sim 1/\langle\sigma\rangle$,  the 
counterterms are higher dimensional. They must however respect the  scale symmetry 
of the lhs of eq.(\ref{tv});  one can then ``restore'' this symmetry ``broken'' by 
working with the  truncation of the rhs expression, by simply  replacing  
$1/\langle\sigma\rangle\ra 1/\sigma$ in their expression. Therefore, the new 
counterterms  of the theory  are suppressed by powers of $\sigma$ and are 
non-polynomial in fields; log-terms in  $\sigma$ are also possible, however (see later).

For example, for $V(\phi)=\lambda\phi^4/4!$, a first counterterm 
is found  by inserting a single internal line of $\tilde\sigma$ in an amplitude,
which brings a factor  ($\epsilon/\langle\sigma\rangle)^2$; if this multiplies  a 
$1/\epsilon^3$ pole  from a three-loop momentum integral it generates a
 $1/\epsilon$ pole and a  corresponding counterterm $\phi^6/\sigma^2$ for 
the 6-point amplitude ($\phi^6$) \cite{Monin}. By the same argument, {\it finite}
quantum corrections appear at two-loops (if due to dynamics of $\sigma$) or even
one-loop (due to scale symmetry alone).

Since the theory is scale invariant and so it has no dimensionful couplings,
diagrams that 
would otherwise  be proportional to masses automatically vanish.
Then the only possibility to construct scale invariant $d=4$ counterterms  that are 
suppressed by powers of $\sigma$ is to involve appropriate
 powers $\phi^n$, $n>4$ and higher derivatives of $\phi$ and $\tilde\sigma$.
  Therefore the new counterterms   are found on dimensional grounds as
\medskip
\bea
\sum_{n, m\geq 0} a_{mn}\, \frac{\partial^{2n} \alpha^{m+4}}{\sigma^{2 n+m}},\quad\alpha=\phi,\sigma.
\eea

\medskip\noindent
where the derivatives act in all possible ways in the numerator.
This includes the dilaton-dilaton scattering $(\partial_\mu\sigma)^4/\sigma^4$
(see $a-$theorem \cite{a-theorem}) which emerges at three-loops.

We see that quantum scale-invariant theories are
 non-renormalizable \cite{Monin}, unlike their counterpart  with $\mu=$constant which is
not quantum scale invariant but is renormalizable (if initial $d=4$ action
was so).  The latter case is recovered in the
 limit of large $\langle\sigma\rangle$, when fluctuations $\tilde\sigma$
decouple, see eq.(\ref{tv}).  This picture also applies to gauge theories.

\subsection{One-loop potential}\label{sub2}

Let us first review the quantum corrections to the potential in a scale 
invariant toy model at one-loop, before going to higher loops.
Consider $L$ below in $d=4$ for a scalar $\phi$ 
\bea
L=\frac{1}{2} (\partial_\mu\phi)^2+ \frac{1}{2} (\partial_\mu\sigma)^2-V(\phi),
\quad
V(\phi)=\frac{\lambda}{4!}\,\phi^4.
\eea

\medskip\noindent
In  $d\!=\!4-2\epsilon$  the potential becomes  $\tilde V(\phi,\sigma)$ 
of eq.(\ref{tildeV}) with $V$ as above, 
so $\phi$ and $\sigma$ 
do interact as dictated by  scale symmetry of analytically continued $L$. 
The  one-loop potential is then
\bea\label{eqV}
V_1&=&\tilde V-\frac{i}{2}\,\int \frac{d^d p}{(2\pi)^d}\,
\Tr\, \ln \big[ p^2-\tilde V_{\alpha\beta}+i\varepsilon\big]
\nonumber\\[-6pt]
&=&\tilde V+\frac{1}{4\kappa} \sum_{s=\phi,\sigma} \tilde M_s^4\,\Big(\,
 \frac{-1}{\epsilon}+\ln\frac{\tilde M_s^2}{c_0}\,\Big),\qquad \kappa=(4\pi)^2.
\eea
where  $c_0\!=\!4\pi e^{3/2-\gamma_E}$.  $\tilde M_s^2$   are field-dependent (masses)$^2$, 
eigenvalues of the second derivatives matrix 
$\tilde V_{\alpha\beta}$, with $\alpha,\beta\!=\!\phi, \sigma$.
One eigenvalue  $\tilde M_\sigma^4\propto\epsilon^2$,  thus it
cannot generate counterterms at one-loop. Then
\bea
V_1&=& \tilde V+
 \mu^{2\epsilon} 
\,\frac{ V_{\phi\phi}^2}{4\kappa} \,\Big\{
\,
-\frac{1}{\epsilon}
+
\Big(\overline \ln \frac{V_{\phi\phi}}{(z\,\sigma)^2}-\frac{1}{2}\Big)\Big\},
\quad {\rm with}\quad
V_{\phi\phi}=\frac12 \lambda\phi^2.
\eea
with $\overline\ln A\equiv \ln A/(4\pi e^{1-\gamma_E})$. 
It is important to note  that the  factor $\mu^{2\epsilon}$
 is a function of $\sigma$  (see eq.(\ref{mu})) 
and {\it maintains scale invariance\footnote{This can also  be relevant if one wanted to define and use
instead a  non-minimal subtraction scheme.} in} $d=4-2\epsilon$. Here we work
in the  minimal subtraction scheme (MS).
Thus the (scale-invariant) counterterm is
\bea
\delta L_1
=-\mu^{2\epsilon} \,
\frac{1}{4!}\,\delta_\lambda^{(1)}\, 
\lambda\,\phi^4
\quad
{\rm with} 
\quad
\delta_{\lambda}^{(1)}\equiv Z_\lambda-1=
\frac{3\,\lambda}{2\kappa\epsilon}.
\eea
Then the  one-loop potential in $d=4$ is 
\bea\label{eq13}
U=V(\phi)+\frac{1}{4\kappa} V_{\phi\phi}^2\,
\Big[
\,\overline\ln \frac{V_{\phi\phi}}{(z\sigma)^2} -\frac{1}{2}
\,\Big]; 
\eea
where we  took the limit $\epsilon\!\ra\! 0$. Note that   $U$ has acquired a 
dependence on $\sigma$  at quantum level (under the log) 
and for this reason its expression is now scale invariant (in $d=4$).

Since  dimensionless $z$ keeps track of the
presence of $\langle\sigma\rangle$, the one-loop beta function $\beta_\lambda^{(1)}$ is found by
demanding the bare coupling $\lambda^B= \mu(\sigma)^{2\epsilon} \,\lambda\,Z_{\lambda}\, Z_\phi^{-2}$
be independent of scaling parameter $z$:
\bea
\frac{d\, \lambda^B}{d\ln z}=0 \quad \Ra \quad
\beta_{\lambda}^{(1)}=\frac{d\lambda }{d\ln z}=\frac{3}{\kappa}\,\lambda^2
\eea
%
which is identical to the result for the case $\mu=$constant\footnote{ 
Unlike in  theories with no dilaton (with explicit SSB  by quantum corrections),
$\beta_\lambda\!=\!0$  is not a  necessary condition for having scale symmetry in our case here \cite{S3,tamarit}
since the spectrum is extended to include a dilaton (spontaneous SSB); thus  a
non-zero $\beta_\lambda$ does not mean the theory cannot be scale invariant.}.
The  Callan-Symanzik (CS) equation for a scale-invariant theory \cite{tamarit}
is easily verified:
\medskip
\bea\label{CS0}
\frac{d U}{d\ln z}=\Big(\frac{\partial}{\partial \ln z}+\beta_{\lambda}^{(1)} 
\frac{\partial}{\partial \lambda}\Big)\, U=O(\lambda^3).
\eea

\medskip\noindent
Consider now  the  limit when 
the dilaton decouples. For this Taylor expand 
the potential for $\sigma=\langle\sigma\rangle+\tilde\sigma$ where 
$\tilde\sigma$ are field fluctuations. The result is
\medskip
\bea\label{cw2}
U=V(\phi)+\frac{1}{4\kappa}\,V_{\phi\phi}^2\Big[\overline\ln \frac{V_{\phi\phi}}{(z\langle\sigma\rangle)^2}
-\frac{1}{2}\Big]+\Delta U
\eea
with
\bea\label{eq17}
\Delta U=\frac{1}{4\kappa}\,V_{\phi\phi}^2\,
\Big(-\frac{\tilde\sigma}{\langle\sigma\rangle}
+\frac{1}{2}\frac{\tilde\sigma^2}{\langle\sigma\rangle^2}+\cdots\Big)
\eea

\medskip\noindent
For $\tilde\sigma\!\ll\! \langle\sigma\rangle$, $\Delta U\!=\!0$ and
we recover the  Coleman-Weinberg  result in a $d\!=\!4$ renormalizable theory obtained
in the usual  DR scheme  of  $\mu\!=$constant($=\!z\langle\sigma\rangle$) 
with explicit SSB (no dilaton). Obviously, the CS equation  is still respected.
One  then proceeds to impose  boundary conditions to define the  quartic
 self-coupling  at $\phi\!=\!\langle\sigma\rangle$:
$\lambda_{\langle\sigma\rangle}\!\!=\!\partial^4U/\partial\phi^4\vert_{\phi=\langle\sigma\rangle}$,
 as usual.

The analysis is very similar if more fields $\phi_j$ are present, of potential $V(\phi_j)$.
The result is found from eqs.(\ref{cw2}), (\ref{eq17})
by replacing $V_{\phi\phi}$ by the eigenvalues of matrix 
$V_{ij}=\partial^2V/\partial\phi_i\partial\phi_j$ and summing over 
them. Again the dilaton  does not contribute counterterms at one-loop, but  enforces 
the scale invariance of $U$ (via $\ln\sigma$). 
 The second term in the CS equation in (\ref{CS0})  is 
now a  sum  over all quartic couplings in $V$. Including fermions and gauge bosons 
is immediate by extending the sum over field dependent masses, with appropriate factors.

\subsection{Two-loop potential}

The two-loop correction to the potential of $\phi$ can be written as
\bea
V_2=V_2^a+V_2^b+V_2^c
\eea

\medskip\noindent
with the diagrams below computed from the background field method\footnote{
We use the approach of \cite{dzp} but without a  classical coupling $\lambda_m\,\phi^2\sigma^2$.}
\bea\label{twoloop}
 V_2^a = \frac{i}{12}\; 
\parbox[c][2em][c]{4.0em}{\includegraphics[scale=0.14]{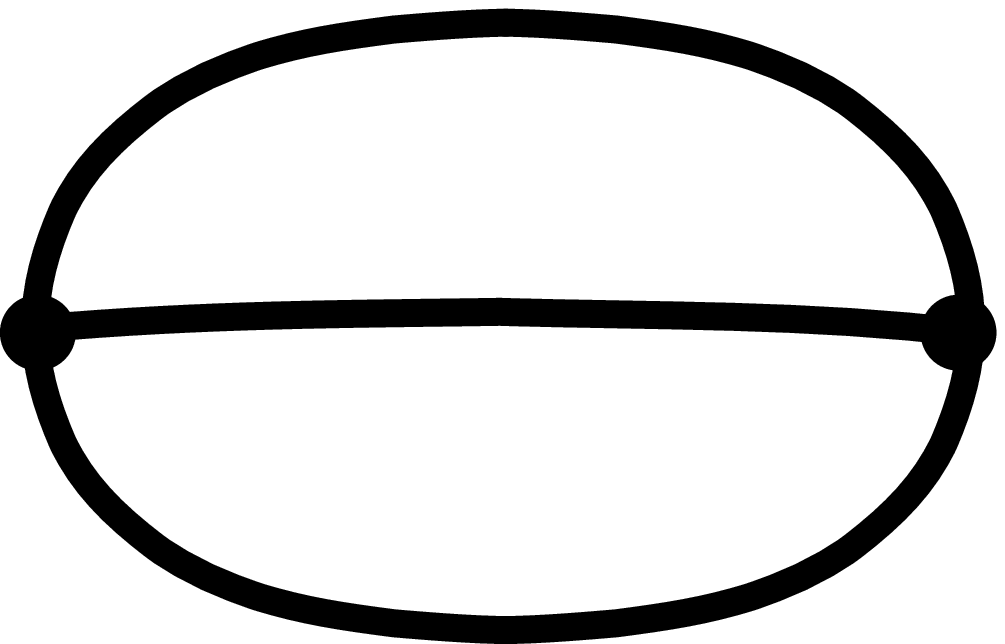}};
\qquad\qquad
V_2^b =
 \frac{i}{8}\; 
 \parbox[c][3em][c]{1.9em}{\includegraphics[scale=0.14]{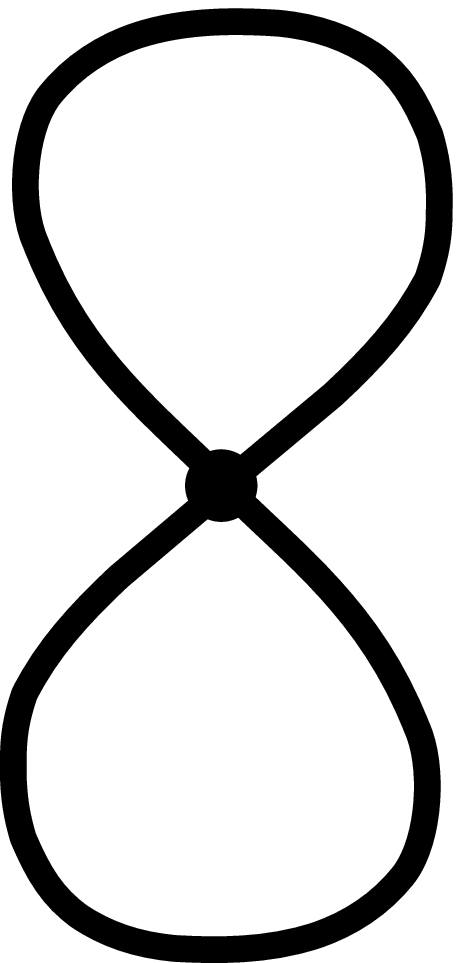}};
\qquad\qquad
V_2^c =  \frac{i}{2}\; 
 \parbox[c][2em][c]{3.2em}{\includegraphics[scale=0.14]{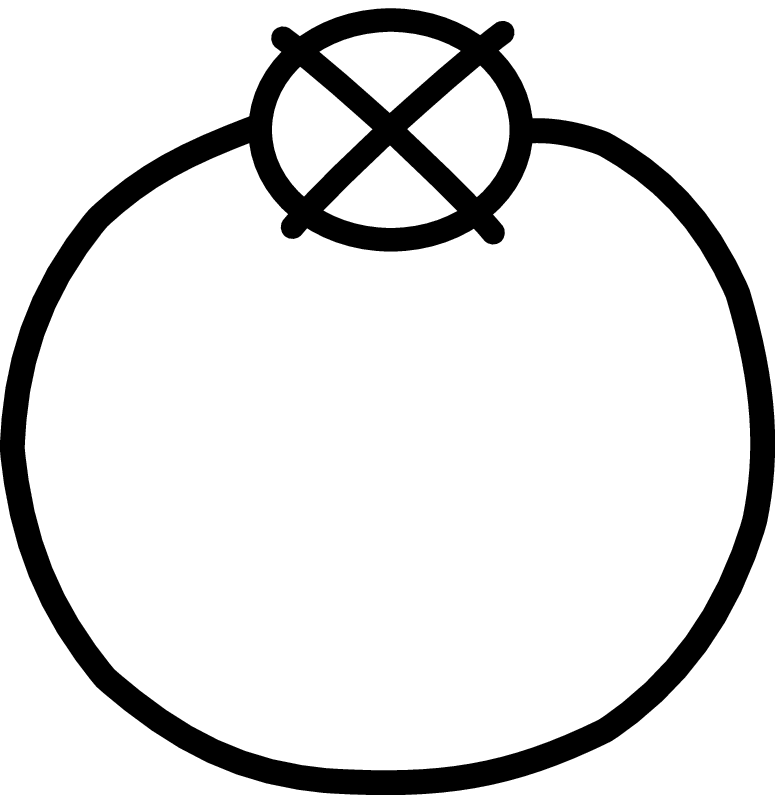}}
\eea

\medskip\noindent
The  vertices and propagators in these diagrams receive evanescent corrections from
 the dilaton field,  as seen from the background field expansion. We Taylor expand
\be
\tilde V(\phi+\delta\phi,\sigma+\delta\sigma)=\tilde V(\phi,\sigma)
+\tilde V_\alpha s_\alpha
+ \frac{1}{2} \tilde V_{\alpha\beta} s_\alpha s_\beta
+\frac{1}{3!} \tilde V_{\alpha\beta\gamma} s_\alpha  s_\beta s_\gamma
+\frac{1}{4!} \tilde V_{\alpha\beta\gamma\rho} s_\alpha s_\beta s_\gamma s_\rho+\cdots
\ee

\medskip\noindent
where $s_\alpha=\delta\phi, \delta\sigma$ are the actual field fluctuations.
The vertices $\tilde V_{\alpha\beta...}=\partial \tilde V/\partial\alpha\partial\beta....$ 
($\alpha, \beta,...\!=\!\phi,\sigma$) contain terms
 proportional to powers of $\epsilon$, e.g.
$\tilde V_{\phi\phi\sigma}=\lambda\, (\phi^2/\sigma) \,\epsilon+\cO(\epsilon^2)$.
The propagators, obtained from the inverse of the matrix 
$(p^2 \delta_{\alpha\beta}-\tilde V_{\alpha\beta})$, also acquire $\epsilon$-dependent
 shifts. We  retain all these  corrections up to  and including $\cO(\epsilon^2)$;  these
can  multiply the poles of the loop integrals ($1/\epsilon^2$ or $1/\epsilon$)
to generate finite  quantum corrections\footnote{
New $1/\epsilon$ poles  from $(\epsilon-{\rm shifts})\times1/\epsilon^2$  
do not emerge here, unless a classical mixing $\phi-\sigma$ exists.}, 
as discussed in Section~\ref{sub1}. Here we shall identify these corrections.
One finds
\smallskip
\bea
V_2&=&
\mu^{2\epsilon}
\frac{\lambda^3\,\phi^4}{32 \kappa^2}
\,
\Big\{ -\frac{3}{\epsilon^2}+\frac{2}{\epsilon}
+\cO(\epsilon^0)
\Big\}.
\eea
with $\mu^{2\epsilon}$ a function of $\sigma$ which {\it maintains the 
scale invariance in} $d=4-2\epsilon$, see eq.(\ref{mu}).
 The counterterm is scale invariant and in the MS scheme is given by
\bea
\delta L_2=
\frac{1}{2} \,
(\partial_\mu\phi)^2\,
\delta_\phi^{(2)}\,
-\mu^{2\epsilon} \frac{1}{4!}
\,\lambda\, \phi^4\,
\delta_{\lambda}^{(2)}\,\, 
\eea
and
\bea\label{d2}
 \delta_{\lambda}^{(2)} =   
\frac{\lambda^2}{\kappa^2}\, \Big(\frac{9}{4\,\epsilon^2}-\frac{3}{2\, \epsilon}  \Big),
\qquad\quad
 \delta_\phi^{(2)}= 
\frac{-\lambda^2}{24 \,\kappa^2 \,\epsilon}.
\eea

\medskip\noindent
From these and with the coefficients $Z_\lambda=1+\delta_\lambda$, $Z_\phi=1+\delta_\phi$ 
and since $\lambda^B=\mu^{2\epsilon} \lambda\,Z_\lambda\,Z_\phi^{-2}$, $d\lambda^B/d(\ln z)=0$,
 one obtains the two-loop corrected beta function
\medskip
\bea
\beta_{\lambda}=\frac{3}{\kappa}\,\lambda^2-\frac{17}{3\,\kappa^2}\, \lambda^3
\eea
$\beta_\lambda$  is identical   to that of the
$\phi^4$  theory  with  $\mu\!=$constant (no dilaton) \cite{S,Panzer,Kleinert}. 
 No new poles (i.e. counterterms)
  are generated at two-loop beyond those of the theory  with $\mu$=constant.

The two-loop potential we find is
\medskip
\bea\label{U}
U&=&\frac{\lambda}{4!}\,\phi^4\, \Big\{\,
1 +
\frac{3\lambda}{2\kappa}\Big(\overline \ln \frac{V_{\phi\phi}}{(z\sigma)^2}
-\frac{1}{2}\,\Big)
+
\frac{3\lambda^2}{4\kappa^2}\,\Big(
4+A_0 -4\,\overline \ln \frac{V_{\phi\phi}}{(z\sigma)^2}
+3\, \overline\ln^2 \frac{V_{\phi\phi}}{(z\sigma)^2}\,\Big)
\nonumber\\[5pt]
&&\qquad\quad+\,
\frac{5\lambda^2}{\kappa^2} \frac{\phi^2}{\sigma^2}
+
\frac{7\lambda^2}{24 \kappa^2}\frac{\phi^4}{\sigma^4}\,\Big\},
 \eea

\medskip\noindent
where\footnote{The Clausen function
 Cl$_2$ is defined as Cl$_2[x]=-\int_0^x d\theta \ln\vert 2 \sin\theta/2\vert$.}
 $A_0=-(8/3)\sqrt{3}\, $Cl$_2(\pi/3)\approx -4.688\cdots$.

Eq.(\ref{U}) is an interesting result. First, $U$ is scale invariant. 
The last two terms in $U$ are new,  finite two-loop corrections in the
form of non-polynomial operators ($\phi^6/\sigma^2$, $\phi^8/\sigma^4$,...)
 and cannot be removed by a different subtraction scheme.
These terms are independent of the dimensionless subtraction parameter $z$
and bring  corrections beyond those obtained for $\mu=$constant (of explicit SSB). 
Their presence is easily understood in the light of the discussion in Section~\ref{sub1}.
The  field-dependent  masses entering the loop calculation, as
 eigenvalues of the second derivative matrix $\tilde V_{\alpha\beta}$, 
contain terms suppressed by $\mu^2\sim\sigma^2$, since the sole 
dependence on $\sigma$ is  $\tilde V\sim \sigma^\epsilon$. This explains the 
presence of positive powers of $\sigma$  only in the denominators  of the non-polynomial terms.
Even the simplest quantum scale invariant
theory is then non-renormalizable (unlike the case with $\mu=$constant 
which is renormalizable but not quantum scale invariant).

The one-loop terms which are  $O(\lambda/\kappa)$ (for $\log\sim 1$)
dominate the new  two-loop non-polynomial terms if 
\medskip
 \bea
\frac{\lambda}{\kappa}\frac{\phi^{n}}{\sigma^n}<1, \qquad   n=2,4.
\eea

\medskip\noindent
The non-polynomial terms can be larger than the  ``standard'' two-loop correction;
they are comparable in size for  $\phi\sim\sigma$.
 Higher loops are expected  to generate more such operators of larger powers and
with new couplings  (if they are counterterms\footnote{This is discussed in the next section.}).
They are relevant if one is interested in the stability of the potential at large field values 
$\phi\sim\langle\sigma\rangle$. The non-polynomial terms vanish in the limit  $\phi\ll \sigma$.

The result in eq.(\ref{U}) can be Taylor expanded about the vev
of  $\sigma$ using  $\sigma=\langle\sigma\rangle+\tilde \sigma$.
Retaining only the leading term  corresponds to decoupling the dilaton. 
Then
\medskip
\bea\label{U2}
U
=\frac{\lambda}{4!}\,\phi^4\, \Big\{
1\!\!\!
&+&\!\!
\frac{3\lambda}{2\kappa}\,\Big(\overline 
\ln \frac{V_{\phi\phi}}{\langle z\,\sigma\rangle^2}
-\frac{1}{2}\Big)
\nonumber\\
&+&\!\!
\frac{3\lambda^2}{4\kappa^2}\,\Big(
4+ A_0\! -4 \,\overline \ln \frac{V_{\phi\phi}}{\langle z\,\sigma\rangle^2}
+3\, \overline\ln^2 \frac{V_{\phi\phi}}{\langle z\,\sigma\rangle^2}\Big)
\Big\} +\cO\Big(\frac{1}{\langle\sigma\rangle}\Big)
\eea

\medskip\noindent
Ignoring $\cO(1/\langle\sigma\rangle)$ terms,
eq.(\ref{U2}) is  the ``standard'' two-loop result obtained for $\mu$=constant 
(no dilaton, explicit SSB) in MS scheme \cite{2loop}, more exactly 
for  $\mu=z\langle\sigma\rangle$.
The difference between eq.(\ref{U}) and eq.(\ref{U2}) is made of  higher dimensional
operators suppressed by  large  $\langle\sigma\rangle$;  these suppressed terms 
are responsible for maintaining manifest scale invariance of  (\ref{U}).

The generic form of the Callan-Symanzik equation is \cite{tamarit}
\medskip
\bea \label{CS}
\Big(
\frac{\partial}{\partial \ln z}
+
\beta_{\lambda_j}\, \frac{\partial}{\partial \lambda_j} 
+
\phi\gamma_{\phi}\,\frac{\partial}{\partial\phi}
+
\sigma\gamma_\sigma\,\frac{\partial}{\partial\sigma}\Big)\, U(\phi_j,\sigma, \lambda_j, z)
=0,
\eea

\medskip\noindent
and we use it to check the result of (\ref{U}). Here\footnote{At two-loop $\gamma_\phi$ is 
$\gamma_\phi^{(2)}=-\lambda^2/(12\kappa^2)$, from eq.(\ref{d2}).}
\bea
\gamma_\phi=\frac{d\ln \phi}{d \ln z}=\frac{-1}{2}\frac{d \ln Z_\phi}{d \ln z}.
\eea

\medskip\noindent
To check eq.(\ref{CS}), first use eq.(\ref{U}) to introduce a decomposition 
$U=V+V^{(1)}+V^{(2)}+V^{(2,n)}$ to denote the tree-level ($V$),  one-loop ($V^{(1)}$),
the ``usual'' two-loop correction with $\mu\ra z\sigma$  ($V^{(2)}$) and finally,
 the  new finite two-loop correction ($V^{(2,n)}$) of the non-polynomial operators 
(the sum of the last two terms in (\ref{U})). Eq.(\ref{CS}) is decomposed into 3 equations:
\medskip
\bea
&& \frac{\partial \, V^{(1)}}{\partial \ln z} \,
 + \,  \beta^{(1)}_{\lambda}\, \frac{\partial V}{\partial \lambda}  = O(\lambda^3) 
\label{cs1}
\\
&& \frac{\partial \, V^{(2)} }{\partial \ln z} 
 + \left( \beta^{(2)}_{\lambda} 
\frac{\partial \;}{\partial \lambda}
+ \gamma_\phi^{(2)} \phi \frac{\partial}{\partial \phi}
+ \gamma_\sigma^{(2)} \sigma \frac{\partial}{\partial \sigma}\right) 
V +  \beta^{(1)}_{\lambda} \,\frac{\partial V^{(1)}}{\partial\lambda}=O(\lambda^4)
\label{cs2}
\\
&&
\frac{\partial\, V^{(2,\text{n})}}{\partial \ln z}=O(\lambda^4),
\label{cs3}
\eea

\medskip\noindent
where $\beta_{\lambda}^{(k)}$, $k=1,2,..$ denote the $k-$loop correction to 
the beta function of $\lambda$  (similar for  $\gamma_{\phi,\sigma}^{(2)}$).
We verified  that eqs.(\ref{cs1})-(\ref{cs3}) are respected. This is a 
consistency check of eq.(\ref{U}).

The Callan-Symanzik equation is also respected in the non-scale-invariant case, 
eq.(\ref{U2}), where  $\mu$=constant ($\mu= z\langle\sigma\rangle$, explicit SSB). 
This is obvious from the above check because $z$ is tracking exactly this scale 
and the non-polynomial terms in (\ref{U}) are $z$-independent\footnote{This 
changes at 3-loops, see $V^{(3,n)}$ in the next section.}.

\medskip
\subsection{Three-loop potential}

In this section we use the three-loop Callan-Symanzik equation for the scalar 
potential to identify the three-loop correction to the  potential without
doing the  diagrammatic calculation. As in the two-loop case, this correction is a 
a sum of two terms $V^{(3)}\!+ V^{(3,n)}.$ $V^{(3)}$ is the ``usual'' three-loop correction
obtained with $\mu\!=$constant (no dilaton) \cite{2loop,Martin}, but with the formal replacement 
$\mu\!\!\ra\! z\,\sigma$;
$V^{(3,n)}$ is a new correction that contains non-polynomial terms.  To find these  
we use the three-loop  counterterms for this theory nicely computed in \cite{Monin}
\medskip
\bea\label{33}
\delta L_3=\frac{1}{2} \,\delta_{\phi}^{(3)}\,(\partial_\mu\phi)^2
-\mu^{2\epsilon}\Big(\,
\frac{1}{4!}\,\delta_{\lambda}^{(3)}\,\lambda\,\phi^4
+
\frac{1}{6}\,\delta_{\lambda_6}^{(3)}\,\lambda_6\, \frac{\phi^6}{\sigma^2}
+
\frac{1}{8}\,\delta_{\lambda_8}^{(3)}\,\lambda_8\, \frac{\phi^8}{\sigma^4}\,
\Big)
\eea

\medskip\noindent
$\delta L_3$ is scale-invariant in $d=4-2\epsilon$ (as it should) 
because  $\mu$ depends on $\sigma$, eq.(\ref{mu}). 
The terms $\phi^6/\sigma^2$ and $\phi^8/\sigma^4$ are expected since
they were present as finite operators at two-loop; also
\medskip
 \bea
\delta_\phi^{(3)}=
-\frac{\lambda^3}{4\kappa^3}
 \Big( \frac{1}{6\epsilon^2}-\frac{1}{12\epsilon}\Big)
\eea

\medskip\noindent
in the MS scheme, giving $\gamma_\phi^{(3)}=\lambda^3/(16\kappa^3)$ and
\medskip
\bea
\delta_{\lambda_6}^{(3)}=\frac{3}{2} \frac{\lambda^4}{\lambda_6\,\kappa^3\,\epsilon},
\qquad
\delta_{\lambda_8}^{(3)}=\frac{275}{864}\frac{\lambda^4}{\lambda_8\,\kappa^3\,\epsilon}.
\eea

\medskip\noindent
With $\lambda_6^B=\mu^{2\epsilon}(\sigma) \lambda_6 Z_{\lambda_6} Z_\phi^{-3} Z_\sigma$, etc, 
and  with $(d/  d\ln z)\, \lambda_6^B=0$, we  find
\bea
\beta_{\lambda_6}&=&
\frac{\lambda^2\,\lambda_6}{2\kappa^2}+\frac{\lambda^3}{\kappa^3}\,\Big( 9\lambda
-\frac{3}{8} \lambda_6\Big)
\nonumber\\
\beta_{\lambda_8}&=&
\frac{2\lambda^2\,\lambda_8}{3\kappa^2}
+\frac{\lambda^3}{4\kappa^3}\,
\Big(\frac{275}{36}\lambda-2 \lambda_8\Big)
\eea

\medskip\noindent
Both beta functions have a two-loop part (hereafter denoted $\beta_{\lambda_{6,8}}^{(2,n)}\sim 1/\kappa^2$)
that is absent if $\lambda_{6,8}=0$  in the classical Lagrangian, which is our case 
here\footnote{Otherwise the terms $\phi^6/\sigma^2$ and $\phi^8/\sigma^4$ would have been 
counterterms  already at two-loop, in eq.(\ref{U}).}; then the three-loop part 
(hereafter $\beta_{\lambda_{6,8}}^{(3,n)}\sim 1/\kappa^3$)  is induced by $\lambda$ alone.
These beta functions  enter in the CS equations in the presence of $\lambda_6$ and
$\lambda_8$, due to their associated counterterms. 
In their presence, eq.(\ref{cs1}) is unaffected, but
eq.(\ref{cs2}) is modified such as  $V$ is now replaced by
\smallskip
\bea\label{newV}
V\ra V+\Delta V,
\qquad \Delta V=
\frac{\lambda_6}{6}\frac{\phi^6}{\sigma^2}+\frac{\lambda_8}{8}\frac{\phi^8}{\sigma^4}.
\eea 

\smallskip\noindent
and  $\beta^{(2,n)}_{\lambda_{6, 8}}$ are also  included in   the first term under the big bracket 
of (\ref{cs2}).   Using these and ``new'' $V$  above, 
one immediately sees that (\ref{cs2}) is verified for non-zero $\lambda_{6,8}$.

Further, there is a CS equation at order $\lambda^4$ for $(V^{(3)}\!+\!V^{(3,n)})$,
which we divide into two CS equations, eqs.(\ref{sss}) and (\ref{v3n}) below. 
One equation is   for the ``usual'' 
correction  $V^{(3)}$  and is identical to that obtained for  
$\mu=$constant ($=z\langle\sigma\rangle$)
\smallskip
\bea\label{sss}
\quad\frac{\partial V^{(3)}}{\partial\ln z}+
\beta_{\lambda}^{(1)}\frac{\partial V^{(2)}}{\partial \lambda}
+
\beta_{\lambda}^{(2)}\frac{\partial V^{(1)}}{\partial \lambda}
+
\beta_{\lambda}^{(3)}\frac{\partial V}{\partial \lambda}
+
\gamma^{(2)}_\phi \,\frac{\partial V^{(1)}}{\partial\ln \phi}
+
\gamma^{(3)}_\phi \,\frac{\partial V}{\partial\ln \phi}
=\cO(\lambda_j^5).
\eea 

\smallskip\noindent
We integrate (\ref{sss}) to find  $V^{(3)}$  up to an unknown ``constant'' of integration
term $\propto \mathcal{Q}$
\smallskip
\be\label{vv3}
V^{(3)}=\frac{\lambda^4\,\phi^4}{\kappa^3}\,
\Big\{\mathcal{ Q}+ 
\Big(\frac{97}{128} +\frac{9}{64} A_0 +\frac{\zeta[3]}{4}\Big)\,
\overline \ln \frac{V_{\phi\phi}}{(z\sigma)^2}
-\frac{31}{96}\overline\ln^2 \frac{V_{\phi\phi}}{(z\sigma)^2} 
+\frac{9}{64} \overline\ln^3\frac{V_{\phi\phi}}{(z\sigma)^2}\Big\}.
\ee

\smallskip\noindent
$\mathcal{Q}$ can be read from the ``usual'' 
 three-loop computation at $\mu=$constant \cite{2loop} in MS scheme:
\bea\label{tre}
\mathcal{Q}\equiv\frac{1}{288}\Big\{
\!\frac{-1673}{8}\! +\frac{9}{4}  A_0 (A_0-4)\! +\frac{34 \pi^4}{15}\! +\! 8\pi^2\ln^2 2
-8\ln^4 2 - 192\, \Li_4\Big[\frac12\Big]\! + 72 \zeta[3]
\Big\}.
\eea

\medskip\noindent
$A_0$ is defined after eq.(\ref{U}), $\Li_4[x]$ is the polylogarithm and 
  $\zeta[x]$ is the Riemann Zeta function.

Finally, there is one last three-loop CS equation, similar to
 (\ref{cs3}), that involves  $V^{(3,n)}$
\smallskip
\bea\label{v3n}
\frac{\partial V^{(3,n)}}{\partial \ln z} 
+\beta_{\lambda_j}^{(1)} \frac{\partial V^{(2,n)}}{\partial\lambda_j}
+\beta_{\lambda_j}^{(3,n)} \frac{\partial V}{\partial\lambda_j}
=\cO(\lambda_j^5), \qquad \lambda_j=\lambda,\lambda_6,\lambda_8.
\eea

\medskip\noindent
where $\beta_\lambda^{(3,n)}$ denotes possible three-loop corrections beyond $\beta_\lambda^{(3)}$.
Eq.(\ref{v3n}) is actually a field-dependent condition. 
As usual $V^{(3,n)}$  only involves  new field operators beyond $V^{(3)}$,  suppressed 
by $\sigma$, e.g.  $\phi^6/\sigma^2$, etc. 
The last term  in the lhs with $\lambda_j\ra\lambda$ would bring a term
$\propto\phi^4$ which cannot be cancelled, being  the only one of this structure.
Then the only way to respect the above field-dependent  condition is that
$\beta_{\lambda}^{(3,n)}=0$. This is  also seen from (\ref{v3n})  in  the  
decoupling limit of large $\langle\sigma\rangle$.
Therefore, the three-loop beta function  in the quantum scale 
invariant effective theory  is just that of the theory with 
$\mu=$constant\footnote{Therefore we have
$\beta_{\lambda}^{(3)}=\lambda^4/\kappa^3\,\big(\,145/8+12\,\zeta[3]\,\big)$  
\cite{S,Panzer,Kleinert}.}$^,$\footnote{This 
is also consistent with $Z_\sigma=1$ at three-loops. A three-loop wavefunction 
correction to $\sigma$  generated by a  coupling $\epsilon\sigma\phi^4$ 
would then be proportional to $\propto\epsilon^2\times (1/\epsilon^2)$,
 so no new poles emerge in this order.}.
We then integrate  eq.(\ref{v3n}) using the replacement 
$\ln z\ra (-1/2)\,\overline\ln(V_{\phi\phi}/(z\,\sigma)^2)$
which fixes the ``constant'' of integration in a  scale invariant way. We find\footnote{
``Constants'' of integration  $\phi^6/\sigma^2$, $\phi^8/\sigma^4$, $\phi^4$
are not allowed, being ``fixed'' in (\ref{33}), ((\ref{vv3}), (\ref{tre}) for $\phi^4$).} 
\medskip
\bea
V^{(3,n)}=\frac{\lambda^3}{2\,\kappa^3}\,\phi^4\,
\Big\{\Big( 27 \lambda-\frac{\lambda_6}{2}\Big) \frac{\phi^2}{8\,\sigma^2}
+\Big(\frac{401\,\lambda}{72}-\lambda_8\Big)\frac{\phi^4}{16\,\sigma^4}
\Big\}\,
\overline \ln \frac{V_{\phi\phi}}{(z\sigma)^2}
\eea

\medskip\noindent
$V^{(3,n)}$ is correct up to a possible additional presence of a scale invariant $z-$independent
three-loop {\it finite} (non-polynomial) term   $(\lambda^4/\kappa^3)\,\phi^{10}/\sigma^6$ 
that cannot  be captured by  the CS differential equation but only in the diagrammatic approach.
In the limit of large field  $\sigma$ and similar to $V^{(2,n)}$ at two-loop,
$V^{(3,n)}\ra 0$, leaving  ``usual'' $V^{(3)}$   as the sole three-loop correction to 
the potential, with only a log-dependence on $\sigma$.

To conclude,  quantum scale invariance demands the presence of  
non-polynomial operators. This symmetry  arranges them in a series 
expansion in powers of $\phi/\sigma$ that contributes to the scalar potential.  
Each of these operators is actually  an  infinite sum of  polynomial  
operators (in fields), after a Taylor   expansion   about  
$\sigma=\langle\sigma\rangle+\tilde\sigma$. $V^{(2,n)}$, $V^{(3,n)}$, $\Delta V$ 
are relevant for  the behaviour of  the potential at large $\phi\sim\sigma$
and are suppressed at $\phi\ll\sigma$.

 \subsection{More operators}

 Having seen the scale invariant non-polynomial operators generated at loop level, 
 it is of interest to see their role if they are included  in the action already
 at classical level, as in
 \bea
 V=\frac{\lambda}{4!}\,\phi^4+ \frac{\lambda_6}{6} \frac{\phi^6}{\sigma^2}+\cdots
 \eea
where we ignore similar higher order terms.
The last term breaks the  enhanced Poincar\'e symmetry ($P_v\!\times\! P_h$) only
mildly, since this symmetry is restored at large $\sigma$. 
In a consistent setup like Brans-Dicke-Jordan theory of gravity, this operator
suppressed by  $\langle\sigma\rangle\sim M_{\rm Planck}$  could mediate gravitational 
interactions of the  two sectors.  Such operator is also generated when going from Jordan to
Einstein frame, after a conformal transformation\footnote{We ignore here the effect 
of $\phi$ on the vev of $\sigma$.}. 

The one-loop computation of the potential  proceeds as before and has three
contributions, all scale invariant. First, there is a one-loop contribution  
similar to that in  eq.(\ref{eq13})  with $V_{\phi\phi}$ replaced  by the (two) 
field-dependent (masses)$^2$ which are  eigenvalues of the matrix of 
 second derivatives  of $V$  above wrt $\phi$ and $\sigma$, then sum over these.

A second contribution to the potential exists.
The two field-dependent masses derived from $\tilde V$ of eq.(\ref{tildeV})
with $V$ as above have a correction $\cO(\epsilon)$ induced by $\lambda_6$; when this 
multiplies $1/\epsilon$ of eq.(\ref{eqV}), it generates a finite correction
$V^{(1,n)}\propto\lambda_6$  already at one-loop
\bea\label{extra}
V^{(1,n)}=\frac{\lambda_6}{6\kappa}\,\phi^4\,
\Big(
4 \lambda\, \frac{\phi^4}{\sigma^4}
+24\lambda_6\,\frac{\phi^{6}}{\sigma^6}
+5\lambda_6\,\frac{\phi^{8}}{\sigma^8}
\Big).
\eea

\medskip\noindent
Finally, there are also  one-loop counterterms,  of the form
$(Z_{\lambda_p}-1)\, \lambda_{p}\,\phi^{p}/(p\,\sigma^{p-4})$, where $p=6,8,10,12$ and
where
$Z_{\lambda_p}=1+\gamma_{\lambda_p}/(\kappa\epsilon)$ and $\gamma_{\lambda_6}=9\lambda$,
$\gamma_{\lambda_8}=56\lambda_6/\lambda_8$, $\gamma_{\lambda_{10}}=20\lambda_6^2/\lambda_{10}$,
$\gamma_{\lambda_{12}}=3\lambda_6^2/\lambda_{12}$.
Therefore the potential has  a third contribution
\medskip
\bea
\Delta V=\sum_{p} \frac{\lambda_p}{p}\frac{\phi^p}{\sigma^{p-4}},  \quad p=6,8,10,12.
\eea
$V^{(1,n)}$ and $\Delta V$ are  similar to  $V^{(2,n)}$, $V^{(3,n)}$, $\Delta V$  found in 
the previous section, except that they are generated at one-loop, due to non-zero $\lambda_6$.
The one-loop beta functions of $\lambda_p$ are
\bea
\beta^{(1)}_{\lambda_{p}}=\frac{2}{\kappa}\lambda_p\,\gamma_{\lambda_p}, 
\eea
with $p$ as above and they vanish if  $\lambda_6=0$.
We checked that the  one-loop CS equation is again verified in the presence of these operators.
For large $\langle\sigma\rangle$, dilaton fluctuations are suppressed and 
the above corrections to the potential vanish, to leave the ``usual'' result (first contribution above), 
obtained in the renormalizable theory with  $\mu$ constant ($=z\langle\sigma\rangle$).
 The generalisation to more operators in the
classical action is immediate.

\subsection{Symmetries, regularisations and mass hierarchy}

From the above examples, we see that a combination of quantum scale invariance 
and enhanced Poincar\'e symmetry \cite{Foot} of the two sectors
can ensure a protection of the mass corrections to $\phi$ against a  quadratic dependence 
 on the  scale of symmetry breaking  $\langle\sigma\rangle$  (the only UV physical scale here).
No term such as $\lambda\,\phi^2\sigma^2=\lambda\langle\sigma\rangle^2\phi^2+\cdots$
was generated at the quantum level in the potential, with $\lambda$ the higgs self-coupling\footnote{
This term is forbidden for large $\langle\sigma\rangle$ by the enhanced Poincar\'e symmetry
(restored in this limit).};
 if present
this would have required the usual SM-like fine-tuning of
$\lambda$. Further, if  one introduces a classical ``mixing'' coupling $\lambda_m$, with a tree-level term
 $\lambda_m\phi^2\sigma^2$ which  would break the enhanced $P_v\times P_h$ symmetry, 
this would require a tuning of $\lambda_m$ (rather than $\lambda$) upon replacing 
$\sigma\ra\langle\sigma\rangle+\tilde \sigma$, in order to keep the  correction to the mass
of $\phi$ under control.  But such tuning of $\lambda_m$ 
is natural and needs to  be done only once at the 
{\it classical} level, since  the beta function  $\beta_{\lambda_m}\sim\lambda_m$ at one-loop
\cite{Allison:2014zya,S1,Ghilen,Foot} and two-loops \cite{dzp}. 
Further, for large 
$\langle\sigma\rangle$ the non-polynomial operators that broke the $P_v\times P_h$ 
symmetry vanish and this symmetry and its ``protective'' role (on $\lambda_m$) are restored. 
Therefore, this protection  remains true in the presence of non-polynomial operators 
e.g. $\lambda_6\not=0$.\footnote{Since we are using  
spontaneous SSB  and a SR scheme,  the conclusions of \cite{Schmaltz} do not apply here.}

The SR scheme used here is based on the DR scheme which 
may be  considered unsuitable to capture the quadratic UV-scale  
dependence of the  scalar (mass)$^2$. It is important to note, however,
that  in  our approach  any scale is generated by fields vevs after spontaneous SSB. The 
 field-dependence (e.g. counterterms, etc) of the  quantum corrected action   is
 not affected by the regularisation and  is actually dictated by  the symmetries of 
the theory (including scale invariance),  which our SR scheme respects (unlike DR). 
Therefore, the dependence of the quantum action on the mass scales (generated by 
these fields vev's) cannot be  affected. The UV behaviour  of the 
 mass of $\phi$ i.e. its dependence on $\langle\sigma\rangle$ (our physical
UV scale), is thus not affected by a regularisation that respected all 
symmetries of the theory\footnote{
To appreciate  the role of  $d\!=\!4$ enhanced Poincar\'e symmetry, consider
a different scale-invariant regularization  that violates the $P_v\times P_h$ symmetry. 
For  $V=\lambda\phi^4/4!$, use a   momentum ``cutoff'' regularisation:
 $k^2\!\leq \!\sigma^2$;  $\sigma$ is a hidden  sector field 
with $\langle\sigma\rangle$  the scale of new physics. 
At one-loop $\Delta V\!\propto\! \int_0^{\sigma^2}\!\!\!\!\!
d^4 k \ln \big(1\!+\!\lambda\phi^2/(2k^2)\big) \!
=\! \lambda\phi^2\sigma^2\!+...$ This term (absent in our case) requires 
the ``usual'' order-by-order fine-tuning of self-coupling $\lambda$.}.

\section{Conclusion}

Following the original idea of Englert et al and using a perturbative
approach, we examined the  quantum implications
of a regularization scheme that preserves the scale invariance  of the classical theory.
To this purpose, we demanded that the analytical continuation of the theory to $d=4-2\epsilon$
preserves the  scale symmetry of the $d=4$ action. This is possible under the additional
presence of a dilaton field ($\sigma$),  the Goldstone mode of scale symmetry breaking. This field
is classically decoupled from the visible sector, following an enhanced Poincar\'e symmetry
of the two sectors, but there are nevertheless quantum effects.

 The scale invariance in
$d=4-2\epsilon$ and the dilaton it demands  have two main effects:

\noindent
\,\,\,\,\,\,\,\,\,a) introduce new  ``evanescent'' interactions  
($\propto\epsilon$) which have quantum consequences;

\noindent
\,\,\,\,\,\,\,\,\,b) generate the subtraction scale 
$\mu\!\sim\!\langle\sigma\rangle$ after spontaneous scale symmetry breaking.

\noindent
As a result, a scale invariant regularisation is re-formulated into an ordinary DR scheme
of $\mu=$constant ($\propto\langle\sigma\rangle$) plus an additional field (dilaton)
with an {\it infinite series} of evanescent couplings to the visible sector.
 When  evanescent interactions multiply the poles of loop integrals, new 
quantum corrections (finite or infinite counterterms) are generated, not present 
in the quantum version of the same  theory regularized with $\mu=$constant (i.e. no dilaton,
 explicit breaking). These corrections, which also include 
$\log$-like terms in the potential (such as  $\ln\sigma$ already at one-loop!),
 are  scale-invariant. They   have effects 
such as transmission of scale symmetry breaking after its spontaneous breaking in the 
dilaton (hidden) sector, or dilaton-dilaton scattering.

The scalar potential was computed at two-loops by direct calculation
and at three loops by integrating its Callan-Symanzik equation. The result is scale invariant. 
It contains new log-like corrections
(in the dilaton $\sigma$) similar to those obtained by naively  replacing $\mu\ra\sigma$
in the result obtained in the ``usual'' DR scheme with $\mu=$constant.
In addition, depending on the details of the classical theory, 
scale invariant non-polynomial effective operators  are also
generated from one- or two-loops onwards, in a series  of the form 
$\phi^4 \times (\phi/\sigma)^{2n}$. These operators are important for large field values 
$\phi\sim\sigma$ and can be comparable to ``standard'' log terms of the loop corrections;
the beta functions of their couplings  were also  computed.

These operators are a generic presence and  can be understood via their 
Taylor series expansion about the scale   $\langle\sigma\rangle\not=0$ of 
spontaneous SSB,  when they become polynomial.   Scale symmetry acts at 
the quantum level as an organising principle that re-sums the polynomial ones.
Therefore, maintaining at the quantum  level the scale symmetry 
of the classical action makes the theory  non-renormalizable.  
In the decoupling limit of   the  dilaton these  operators 
vanish and  one recovers  the quantum  result  of  a renormalizable  
theory with  explicit  SSB (if classical theory  was renormalizable).

The role of the quantum scale symmetry and enhanced Poincar\'e symmetry in 
protecting a mass hierarchy $m_{\phi}^2\ll \langle\sigma\rangle^2$
was   reviewed. This protection  cannot be affected by working in a
 regularisation ultimately based on a  DR  scheme,  because all scales and thus hierarchy  
thereof are generated by vev's of the fields present in the  quantum corrected action (after 
spontaneous SSB);  its counterterms i.e.  fields dependence are dictated by 
the symmetries of the theory (including scale symmetry), that our regularisation 
respects (unlike DR), hence the aforementioned protection. This remains true  in the presence of
the non-polynomial terms (i.e. despite non-renormalizability)
 since at large $\langle\sigma\rangle$ the enhanced Poincar\'e symmetry is restored.
The study can be extended to gauge theories.

\bigskip

\bigskip\medskip
\noindent
{\bf Acknowledgements: } We thank  P. Olszewski for a discussion on the two-loop case.

\end{document}